\documentclass{article}
\usepackage{amssymb}
\usepackage{amsmath}

\def\abstract#1{\vskip 7mm 
        \begin{center}{\large Abstract}\par \smallskip
                \begin{minipage}[c]{12cm}
                        \small #1
                \end{minipage}
        \end{center}
}
\def\title#1{\begin{center}{\Large\bf #1}\end{center}}
\def\author#1{\vskip 5mm \begin{center}{#1}\end{center}}
\def\address#1{\begin{center}{\it #1}\end{center}}
\makeatletter
\@ifundefined{lesssim}{}{}
\@ifundefined{gtrsim}{}{}
\def\vereq#1#2{\lower3pt\vbox{\baselineskip1.5pt \lineskip1.5pt
\ialign{$\m@th#1\hfill##\hfil$\crcr#2\crcr\sim\crcr}}}
\makeatother

\newcommand{\beq}{\begin{equation}}
\newcommand{\eeq}{\end{equation}}
\newcommand{\bea}{\begin{eqnarray}}
\newcommand{\eea}{\end{eqnarray}}

\newcommand{\e}{\mbox{e}}
\renewcommand{\d}{\mbox{d}}

\renewcommand{\l}{\lambda}

\renewcommand{\a}{\alpha}

\newcommand{\ep}{\varepsilon}

\newcommand{\del}{\delta}

\newcommand{\dg}{\dagger}

\newcommand{\ra}{\rangle}
\newcommand{\la}{\langle}
\newcommand{\prt}{\partial}

\newcommand{\equ}{\!=\!}

\newcommand{\cD}{{\cal D}}

\newcommand{\hH}{{\hat{H}}}

\newcommand{\sla}{\sqrt{\l}}

\newcommand{\vac}{|0\ra}
\newcommand{\cav}{\la 0 |}
\newcommand{\dll}{\frac{dl}{l}}

\begin{document}

% 4 pages, proceedings of the workshop JGRG 17 (Nagoya, Japan, December 2007)
% J. Ambjorn, R. Loll, Y. Watabiki, W. Westra, S. Zohren
\title{%
  Topology change in causal quantum gravity 
}
\author{%
  J. Ambj{\o}rn$^{1,2}$, R. Loll$^2$, W. Watabiki$^3$, W. Westra$^4$ and S. Zohren$^{5}$
}
\address{%
  $^1$
The Niels Bohr Institute, Copenhagen University, 
Blegdamsvej 17, DK-2100 Copenhagen \O, Denmark\\
$^2$
Theoretical Physics Institute, Utrecht University, 
Leuvenlaan 4, NL-3584 CE Utrecht, The Netherlands\\
  $^3$  
Department of Physics, Tokyo Institute of Technology, 
Oh-okayama, Meguro, Tokyo 152-8551, Japan
$^4$
  Department of Physics, University of Iceland, 
Dunhaga 3, 107 Reykjavik, Iceland\\
$^5$
Blackett Laboratory, Imperial College, 
London SW7 2AZ, U.K. and\\
 Department of Physics, Ochanomizu University, Otsuka, Bunkyo-ku, 
Tokyo 112-8610, Japan 
}

\abstract{
 The role of topology change in a fundamental theory of quantum gravity is still a matter of debate. However, when regarding string theory as two-dimensional quantum gravity, topological fluctuations are essential. Here we present a third quantization of two-dimensional surfaces based on the method of causal dynamical triangulation (CDT). Formally, our construction is similar to the $c = 0$ non-critical string field theory developed by Ishibashi, Kawai and others, but physically it is quite distinct. Unlike in non-critical string theory the topology change of spatial slices is well controlled and regulated by Newton's constant. 
 %%Our main technical result is an iterative algorithm to analytically compute diagrams of any spatial and space-time topology.
 }

\section{Causal quantum gravity, topology change and Euclidean quantum gravity}\label{cdt}

Why do we study two-dimensional quantum gravity? Firstly, one can test quantisation procedures for gravity in a simple setting.  Secondly, it has long been known that string theory can be viewed as two-dimensional quantum gravity coupled to matter fields.  This particular view of string theory spawned the development  of  the dynamical triangulation (DT) approach to quantum gravity.  This method is particularly powerful in two dimensions, since exact nonperturbative solutions can be obtained by loop equations or matrix models.

In the nineties the DT approach was invoked in an attempt to nonperturbatively define four-dimensional quantum gravity through computer simulations \cite{4ddt}. The results were not satisfactory however since no suitable semiclassical limit was found. 

To improve this state of affairs the method of \emph{causal} dynamical triangulation (CDT) was developed \cite{98}. Contrary to the aforementioned applications of DT, CDT incorporates some essential Lorentzian features. Recent computer simulations indicate that in four dimensions CDT does lead to a sensible classical limit \cite{4d}, unlike in earlier attempts employing DT.

To better understand the relation between the Euclidean (DT) and the causal (CDT) approach we study a generalisation of the 2d CDT model \cite{cap,cdtsft}. Let us start  with a discussion of the original 2d CDT model as introduced in \cite{98}. 

A natural amplitude in CDT is the so-called proper-time propagator. This amplitude is computed by a functional integral over all ``causal'' geometries with topology $S^1 \times [0,1]$. It computes the transition amplitude between an initial and a final boundary, where all points on the initial boundary are separated a geodesic distance $t$ from the final boundary. Here the term causal geometry 
refers to Euclidean geometries that can be obtained from Lorentzian geometries through a Wick rotation defined in the discrete formalism of CDT.
This restriction requires spatial sections of the geometries to be a single $S^1$ and not change as a function of time. Formally, the proper-time propagator is given by the following equation
\beq\label{2.a0}
G_\l (x,y;t) =
\int \cD [g_{\mu\nu}] \; e^{-S[g_{\mu\nu}]},\quad S[g_{\mu\nu}] = \l \int \d^2 \xi  \sqrt{\det g_{\mu\nu}(\xi)} +
x \oint \d l_1 + y\oint \d l_2,
\eeq
where $\l$ is the cosmological constant, $x$ and $y$ are the boundary cosmological constants and $g_{\mu \nu}$ is the causal world sheet metric.
By taking the continuum limit of a discrete iteration equation it can be shown that the proper-time propagator satisfies the equation 
\beq\label{2.n1}
\frac{\prt}{\prt t} G_\l(x,y;t) = - 
\frac{\prt}{\prt x} \Big[(x^2-\l) G_\l(x,y;t)\Big],
\eeq
which can be solved in a straightforward manner to obtain $G_\l(x,y;t)$.
For some purposes it can be more convenient to study correlators $G_\l (l_1,l_2;t)$ where the lengths of the boundaries are fixed rather than the boundary cosmological constants.
Since the lengths of the boundaries are conjugate to the corresponding boundary cosmological constants, the different propagators are related by Laplace transformations,
\beq\label{2.a5}
G_\l(x,y;t)= \int_0^\infty \d l_2 \int_0^\infty \d l_1\;
G_\l(l_1,l_2;t) \;\e^{-xl_1-yl_2}.
\eeq
Strictly speaking it is not possible to define a disc function for a Lorentzian theory of two-dimensional quantum gravity, assuming that the disc boundary represents an instance of constant time.
The reason is that it is impossible to cover the disc with an everywhere nondegenerate Lorentzian metric. This is however possible if one excises one point. 
Consequently one can define the CDT disc function by the ensemble of punctured discs which is given by 
\beq\label{2x.1}
W_\l(x) = \int_0^\infty \d t \; G_\l (x, l_2=0;t) = \frac{1}{x+\sla}.
\eeq
Starting from the discrete setup one can now also include the possibility of the \textit{spatial} topology to change as a function of proper time $t$ keeping the \textit{space-time} topology fixed to be $S_1\times[0,1]$. In \cite{98} it was shown that the corresponding propagator is given by the partial differential equation
\beq\label{2.53}
 a^{\ep}\frac{\prt}{\prt t} G_{\l,g}(x,y;t) = 
- \frac{\prt}{\prt x} \Big[\Big(a(x^2-\l)+2 g\, a^{\eta-1} W_{\l,g}(x)\Big) 
G_{\l,g}(x,y;t)\Big].
\eeq
Here $a$ is a ultraviolet cutoff, $\eta$ and $\ep$ are the scaling exponents of the regularized disc function and time respectively, and $g$ is a coupling constant assigned to each splitting of the spatial universe. In \cite{98} it was shown that if the coupling constant does not scale, there are only two possible scaling relations:
\bea
 (i) \quad W_{reg} &\xrightarrow[a\to 0]{}& a^{\eta}\, W_\l(x),~~~~\eta < 0, \nonumber\\
t_{reg} &\xrightarrow[a\to 0]{}&  t/a^\ep,~~~~\ep =1, \label{2.51a} \nonumber\\
(ii) \quad W_{reg} &\xrightarrow[a\to 0]{}& {\rm const.} +a^{\eta}\, W_\l(x), 
~~~~\eta=3/2 \nonumber\\
t_{reg}& \xrightarrow[a\to 0]{}&  t/a^\ep,~~~~~\ep=1/2.\nonumber
\eea
The first possibility $(i)$ corresponds to the scaling of causal quantum gravity for $\eta=-1$. 
Inserting this scaling relation into \eqref{2.53} implies that $g$ must be set to zero and one recovers \eqref{2.n1} in which no spatial topology changes are allowed.

For the scaling $(ii)$ one recovers 2d Euclidean quantum gravity as defined through Liouville theory or matrix models. In this case the kinetic term is subdominant and the dynamics is purely governed by the splitting of spatial universes, i.e.
\beq
\frac{\prt}{\prt t} G^e_\l(x,y;t) = - 
\frac{\prt}{\prt x} \Big[2 g  \, W^e_{\l,g}(x) G^e_\l(x,y;t)\Big].\label{EUPDE}
\eeq
It is possible to show that in this continuum limit there is a baby universe at every point in the quantum geometry. 
One can see this by contracting the final boundary of the propagator. After contraction the propagator reduces to the disc function with a marked point that can be located anywhere in the bulk, i.e.
\beq
\frac{\prt W^e_{\l,g}(x) }{\prt\l}= \int_0^\infty \d t \; G^e_{\l,g} (x, l_2=0;t).
\eeq
Inserting this into \eqref{EUPDE} and absorbing the dimensionless factor $2g$ in the cosmological constant, one obtains the disc function of 2d Euclidean quantum gravity
\beq
W^e_\l(x) = \left(x-\frac{1}{2} \sla \right) \sqrt{x+\sla}.
\eeq

\section{Taming the topology changes}
In the previous section we showed how starting from 2d CDT one can obtain 2d Euclidean quantum gravity when allowing for spatial topology changes. Under the scaling relations $(i)$ and $(ii)$ discussed above, there was only the possibility of either zero or infinite numbers of spatial topology changes. However, in \cite{cap} it was shown that there exists a unique renormalization of the coupling constant that leads to a well defined double scaling limit
\beq\label{2.54}
g = g_s a^3.
\eeq
In this scaling limit spatial topology changes are included in a controlled manner. The partial differential equation for the propagator then reads
\beq\label{2.55}	
 \frac{\prt}{\prt t} G_{\l,g_s}(x,y;t) = 
- \frac{\prt}{\prt x} \Big[\Big((x^2-\l)+2 g_s\;W_{\l,g_s}(x)\Big) 
G_{\l,g_s}(x,y;t)\Big].
\eeq
Interestingly, the model described by \eqref{2.55} can be solved to all orders in the coupling constant \cite{cap}. In particular, one obtains for the disc function \cite{cap} that
\beq\label{3.9}
W_{\l,g_s}(x) = \frac{-(x^2-\l)+(x-\a)\sqrt{(x+\a)^2-2g_s/\a}}{2\,g_s}, ~~~\a = u\sla, ~~~u^3-u+\dfrac{g_s}{\l^{3/2}}=0.
\eeq
For $g_s=0$ one recovers the disc function of the pure CDT model without any spatial topology changes, however, as shown in \cite{cap}, it is not possible to obtain the disc function of Euclidean quantum gravity as an analytic continuation in $g_s$.

It is interesting to give a gravitational interpretation to the coupling constant $g_s$. As was mentioned above, the disc function of a Lorentzian theory of 2d quantum gravity needs one point of the manifold to be excised. Since each baby universe that splits off is essentially a disc function, a surface with $N$ baby universes contains $N$ punctures. Because of the Gauss-Bonnet theorem each puncture is associated with a factor of one inverse Newton constant $1/G_N$. Hence, we can make the identification $g_0(a)\equ e^{-1/G_N(a)}$, where $G_N(a)$ denotes the ``bare'' gravitational coupling constant. One can introduce a {\it renormalized} gravitational coupling constant by
\beq\label{5.4}
\dfrac{1}{G_N^{ren}} = \dfrac{1}{G_N(a)}+\dfrac{3}{2}\ln \l a^2,
\eeq
which leads to the identification $e^{1/G_N^{ren}}\equ g_s/\l^{3/2}$. The corresponding scaling limit of 2d Euclidean quantum gravity reads
\beq\label{5.6}
\dfrac{1}{G_N^{ren}} = \dfrac{1}{G_N(a)}+\dfrac{5}{4}\ln \l a^2.
\eeq

\section{A string field theory for causal and Euclidean quantum gravity}

In string field theories (SFT) one defines operators that can create and annihilate strings.
From the $2d$ quantum gravity point of view we thus have a 
third quantization of gravity, where one-dimensional universes can 
be created and annihilated. Such a formalism was
developed in \cite{kawai} for non-critical strings, i.e. 2d Euclidean quantum 
gravity and recently in \cite{cdtsft} as a third quantization 
for CDT reproducing the results of the previous section.

The starting point is the assumption of a vacuum from %%
which universes can be created. We denote this state by $\vac$ and
define creation and annihilation operators:
\beq\label{s1} 
[\Psi(l),\Psi^\dg(l')]=l\del(l-l'),~~~\Psi(l)\vac = \cav \Psi^\dg(l) =0. 
\eeq
The Hamiltonian for the CDT SFT is given by \cite{cdtsft}
\bea\label{s8}
&&\hH = \hH_0 - g_s \int \d l_1 \int \d l_2 \Psi^\dg(l_1)\Psi^\dg(l_2)\Psi(l_1+l_2)\nonumber
\\ &&~~~~~~~~~~\, - \a g_s \int \d l_1 \int \d l_2 \Psi^\dg(l_1+l_2)\Psi(l_2)\Psi(l_1)
-\int \dll \; \rho(l) \Psi(l), 
\eea
The first term is the ``second-quantized'' Hamiltonian of the pure CDT model,
\beq\label{s6}
\hH_0 = \int \dll \; \Psi^\dg (l) H_0(l) \Psi(l),\quad H_0(l) =- l \frac{\prt^2}{\prt l^2}+\l l.
\eeq
%where the CDT Hamiltonian $H_0(l)$ can be read of from \eqref{2.n1} using \eqref{2.a5} and
%\beq\label{s3}
%G_\l(l_1,l_2;t) =\frac{1}{l_2} \la l_2| \e^{-t H_0(l) } |l_1\ra.
%\eeq
The third term corresponds to the splitting of strings with the assigned coupling $g_s$ and the fourth term to the joining of strings. The last term, the tadpole, is responsible for the termination of a string into the vacuum and is simply given by $\rho(l)=\delta(l)$, meaning that only strings of length zero can be terminated. 

The disc function of the model can be written in the string field theory language as follows,
\beq\label{s9}
W_{\l,g_s}(l) =\lim_{t\to \infty}W_{\l,g_s}(l,t) = 
\lim_{t\to \infty} \cav \,\e^{-t \hH} \Psi^\dg(l)\vac.
\eeq
In SFT one derives the amplitudes by solving the so-called Dyson-Schwinger (DS) equations,
\beq\label{s10}
0= \lim_{t\to \infty}\frac{\prt }{\prt t}W_{\l,g_s}(l,t) = 
\lim_{t\to \infty} \cav e^{-t \hH }[\hH, \Psi^\dg(l)]\vac.
\eeq
These equations express the fact that the solution should be slowly varying in time for $t \rightarrow \infty$.
In the limit where the joining of strings is forbidden ($\a \rightarrow 0$), the DS equation \eqref{s10} leads to a closed equation for the disc function,
%Upon evaluation of the commutator using \eqref{s1} the following DS equation for the disc function is obtained, 
\beq\label{s11}
\frac{\prt}{\prt x}\left((x^2-\l)W_{\l,g_s}(x)  + g_s W_{\l,g_s}^2(x)\right) = 1.
\eeq
The solution of equation \eqref{s11} is again given by \eqref{3.9}. This shows that the diagrammatic techniques of \cite{cap} are equivalent to the string field theory techniques of \cite{cdtsft}.
For finite $\a$ the DS equations become considerably more complicated as they cannot be written in closed form. 
To evaluate the higher-genus disc functions, say, one also requires knowledge of the higher-loop correlators.

\section{Discussion}

In this contribution we recalled that the loop-loop correlator used in $c=0$ non-critical 
string theory can be obtained by extending the formalism of CDT by allowing the 
topology of spatial slices to fluctuate. In the 
non-critical string theory these spatial topology fluctuations dominate the dynamics completely. 
It was seen that by introducing a coupling constant and a suitable double-scaling limit one can obtain a $2d$ quantum gravity theory where the changes of spatial topology are well controlled \cite{cap}. The amplitudes of this theory have been computed to all orders in the coupling constant. Evaluation can be done both by diagrammatic techniques \cite{cap} or by a string field theory \cite{cdtsft}. Within the string field theory it is in principle possible to compute diagrams of arbitrary space-time topology. When introducing the merging process of strings in the Hamiltonian, one can iteratively solve the corresponding Dyson-Schwinger equations.

\textbf{Acknowledgments:} J.A., R.L. W.W. and S.Z. acknowledge support by
ENRAGE (European Network on
Random Geometry), a Marie Curie Research Training Network in the
European Community's Sixth Framework Programme, network contract
MRTN-CT-2004-005616. R.L. acknowledges 
support by the Netherlands 
Organisation for Scientific Research (NWO) under their VICI 
program. S.Z. thanks Professor A. Sugamoto for discussion 
and the JSPS Short Term  
Award Program for financial support.

% To compare with the non-critical string field theory of Kawai et al. \cite{kawai} and to make the gravitational interpretation of our coupling constant more apparent % one can do the following rescaling, 
% \beq
% \bar{\Psi} =\frac{1}{g_s} \, \Psi,~~~~\bar{\Psi}^\dag = g_s \, \Psi^\dag.
% \eeq
% The CDT SFT Hamiltonion now reads
% \bea
% &&\hat{\bar{H}} =  \int\dll \; \bar{\Psi}^\dg (l) H_0(l) \bar{\Psi}(l) - \int dl_1 \int dl_2 \bar{\Psi}^\dg(l_1)\bar{\Psi}^\dg(l_2)\bar{\Psi}(l_1+l_2)\nonumber
% \\ &&~~~~~~~~\, - \a g_s^2 \int dl_1 \int dl_2 \bar{\Psi}^\dg(l_1+l_2)\bar{\Psi}(l_2)\bar{\Psi}(l_1)
% -\int \dll \; \bar{\rho}(l) \bar{\Psi}(l), 
% \eea
%
% \bea
% \begin{tabular}{llll}
% $(i)$&CDT SFT & $H_0(l) =- l \frac{\prt^2}{\prt l^2}+\l l$ & $\bar{\rho}(l)= g_s\, \delta(l)$\\
% $(ii)$&Euclidean SFT& $H_0(l)=0$  & $\bar{\rho}(l)=$  \nonumber
% \end{tabular}
%\eea

\end{document}